\input{psfig.sty}
\documentclass[12pt,preprint]{aastex}
\usepackage{amssymb}

\newcommand\flux{erg sec$^{-1}$cm$^{-2}$}

\newcommand\cha{{\it Chandra} }

\begin{document}
\title{Spectral variability of the nucleus of M33 in a \cha/ACIS observation.}
\author{ V. La Parola\altaffilmark{1}
, F. Damiani\altaffilmark{1}, 
G. Fabbiano\altaffilmark{2}, G. Peres\altaffilmark{3}} 
\affil{$^{1}$ INAF - Osservatorio Astronomico di Palermo ``G.S. Vaiana'', Piazza del 
Parlamento 1, 90134 Palermo, Italy; }
\affil{$^{2}$ Harvard Smithsonian Center for Astrophysics, 60 Garden St.,
02138 Cambridge, MA} 
\affil{$^{3}$ DSFA, Sezione di Astronomia- Universit\'a di Palermo, Piazza del 
Parlamento 1, 90134 Palermo, Italy; }
\altaffiltext{2}{e-mail: laparola@oapa.astropa.unipa.it}

\begin{abstract}
We have analyzed a  90 ksec long observation of the bright nucleus of 
M33 made with \cha/ACIS. We detected low-amplitude ($\sim10\%$) highly
significant variability on timescales of $\sim5000$ sec. We also find associated
spectral variability. The two main spectral components (a power-law with
$\Gamma\sim2$ and a multicolor disk -MCD- with kT$\sim0.9-1.2$ keV) vary in 
relative flux. The MCD temperature
also increases with increasing MCD flux. The pattern of variability is
reminiscent of (but not identical to) galactic black hole binaries. A $\sim 15$
M$_{\odot}$ accreting black hole may explain this source.
\end{abstract}
\keywords{Galaxies, individuals  (M33) --- galaxies: nuclei --- X-rays:
individual (M33 X-8)} 

\today

\section{INTRODUCTION}
M33 is a nearby (795 kpc, \citealp{vanden}) Sc galaxy. Its nucleus is the 
brightest X-ray source in the Local Group: with a luminosity of $\sim 10^{39}$ 
erg/sec, it accounts for $\sim 70\%$ of the total luminosity of the galaxy in 
the 0.15-4.5 keV band \citep{trin}. The nature of this source (M33-X8 in
\citealp{trin}) is still unclear. Because of its position, coincident with the 
optical center of the galaxy, it has been suggested that X8 could be a
low-luminosity AGN (see, e.g. \citealp{market}), but this interpretation has
two main drawbacks: no activity has been found in other energy bands and an
upper limit of only $1500 M_{\odot}$ has been put on the mass of the central
black hole \citep{gebh}. \citet{dubus} have found an apparent periodicity of
$\sim 106$ days in a set of ROSAT data with a six years coverage, that they
ascribe to a ``superorbital'' modulation in a binary system with a giant star
orbiting around a black hole. However, this
finding has not been confirmed by the subsequent observations (see e.g. 
\citealp{parmar}), that failed to match the phase prediction. Today, the most 
favoured hypothesis is that M33-X8 is an 
intermediate (or even stellar) mass black hole, likely in a binary system. 
Recent studies with ASCA and BeppoSAX data \citep{takano,parmar} agree in 
describing its X-ray energy spectrum with a power-law with a high energy 
exponential cutoff with $\sim 2$ keV $e$-folding energy, or with a multicolor 
disk black-body (MCD), both consistent with the spectra of Galactic black-hole 
candidates and of ultra luminous sources (ULXs, see. e.g., \citealp{maki}).

In this work we examine a Chandra/ACIS-S observation of the nucleus of M33,
and we discuss its variability and spectral characteristics.

\section{THE OBSERVATION AND DATA REDUCTION}
\label{data}
M33 was observed with Chandra/ACIS on July 6th 2001 (Obs Id: 2023), with an exposure time of 
91.8 ksec. The observation was pointed
to the star formation region NGC604, with the aimpoint on ACIS-I, therefore
the nucleus of the galaxy is $\sim 12'$ off-axis and falls on ACIS-S3. 
Though the image of the nucleus is highly degraded, this proves to be an 
advantage in terms of spectral and temporal analysis, as the photons from the 
nuclear source are not affected by pile-up effects, a major problem if the
source were on-axis.

Due to the degradation of the point spread function (PSF) with increasing 
off-axis angle, the source appears distorted and elongated with an elliptical
shape. The source counts were thus extracted from an 
elliptical region centered on the position of M33-X8 as determined from 
ROSAT/HRI data by \citet{schul}, i.e. RA=$01^h33^m50.92^s$, 
Dec=$30^{\circ}39'36.7''$ (J2000). The axes of the ellipse were determined 
examining the radial profiles along the two main direction of the PSF and 
chosing the radii that enclose 99\% of the emission in each direction, 
namely 40 and 25 arcsec, with a major axis position angle of $53^{\circ}$. The 
background was extracted from a concentric
elliptical annular region with 110 and 70 arcsec major and minor semi-axes
respectively, excluding the portions of the annulus falling outside the chip.

The data analysis was performed using different
tools in the CIAO 2.1 software package, the imaging application DS9 v.2.1,
the timing analysis software package Xronos v.5.18 and the spectral analysis
software XSPEC v11.2.

\begin{figure}[t]
\centerline{\psfig{figure=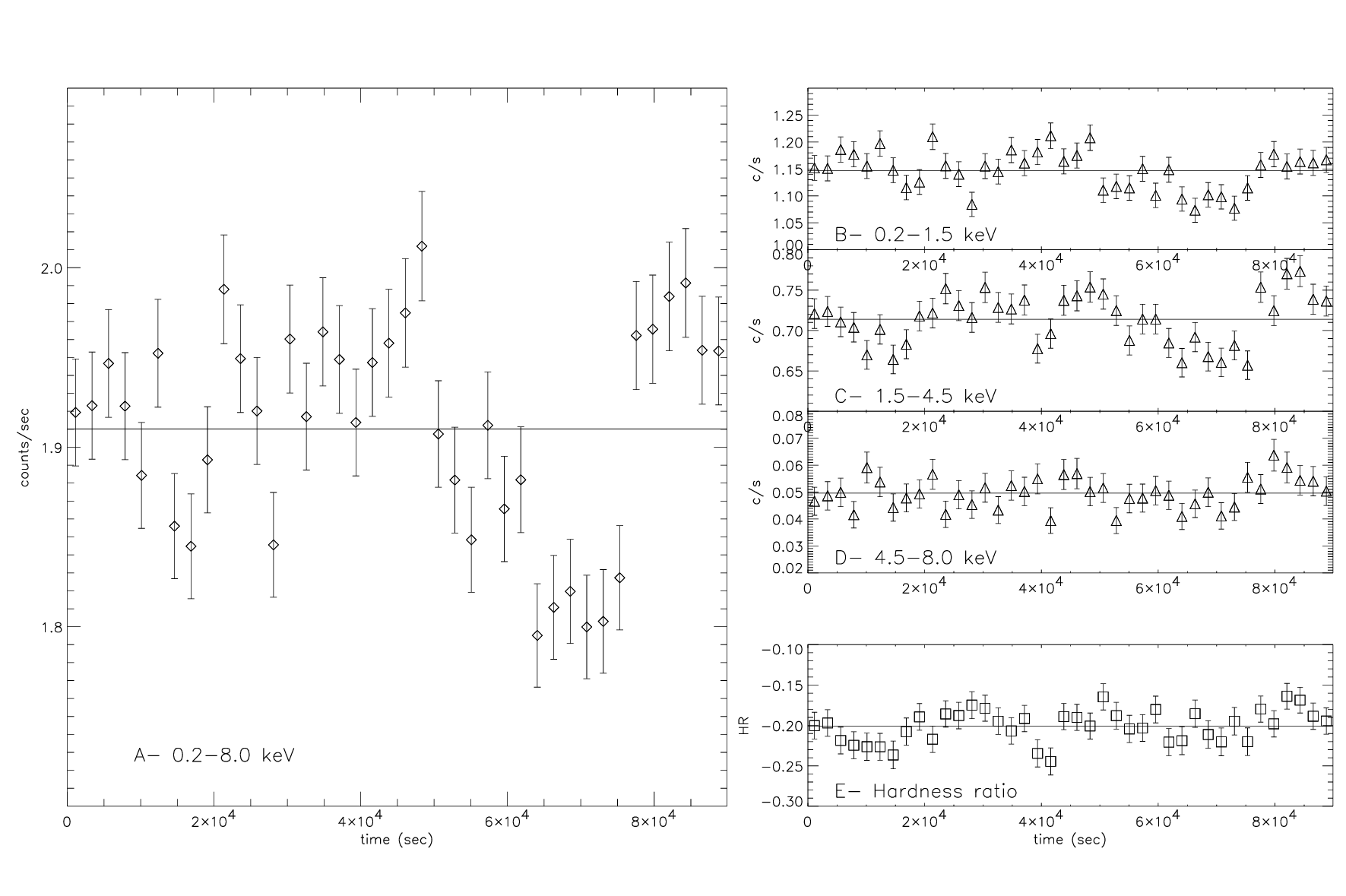,width=18cm,angle=0}}
\caption{Background subtracted light curves of M33-X8. Time is from the
beginning of the observation. Panel A: total \cha 
band (0.2-8.0 keV). 
Panel B: light curve in the 0.2-1.5 keV energy band. Panel C: light curve 
in the 1.5-4.5 keV energy band ; Panel D: light curve in the 4.5-8.0 keV 
energy band. In all panels the
average value is marked as a solid line. Panel E: hardness ratio calculated 
using the 1.5-8.0 keV and 0.2-1.5 keV energy bands. Bins are 2250 s long.}
\label{ltc}
\end{figure}

\section{TEMPORAL ANALYSIS}
\label{time}
The observation is continuous with a time coverage of 91.8 ksec, an average count 
rate of $\sim 1.9$ counts/sec in the 0.2-8.0 keV band and a time resolution of 
3.24 s, that allows for a detailed temporal analysis with moderate resolution. The 
light-curve of M33-X8 has been binned in 40 intervals, each 2250 s long. The
source is evidently variable: the light curves show a significant variability of
more than 10\% from maximum to minimum (Figure~\ref{ltc}- Panel A).
Panels B, C, D in Figure~\ref{ltc} show the light-curve in three energy
sub-ranges (0.2-1.5 keV, 1.5-4.5 keV, 4.5-8.0 keV), chosen to search for
energy-dependent variability: the soft band samples the spectral region that
hosts most of the absorption edges and spectral lines, the medium and hard 
bands cover the continuum and their ratio measures the slope of the continuum
itself. We applied to each band (including the full band) various tests: the 
Kolmogorov-Smirnof test \citep[see, e.g.,][]{cono}, the $\chi^2$ test, the 
Collura test (a phase averaged $\chi^2$ test plus an interpretation with a 
pulse model, \citealp{coll}). These tests (see Table~\ref{tflux}) confirm the 
presence of variability with high statistical significance in the full band and 
in the soft and medium bands. The Collura test indicates a
variability timescale of $\sim 5000$ sec for the total, soft and medium bands.

\begin{figure}[!t]
\centerline{\psfig{figure=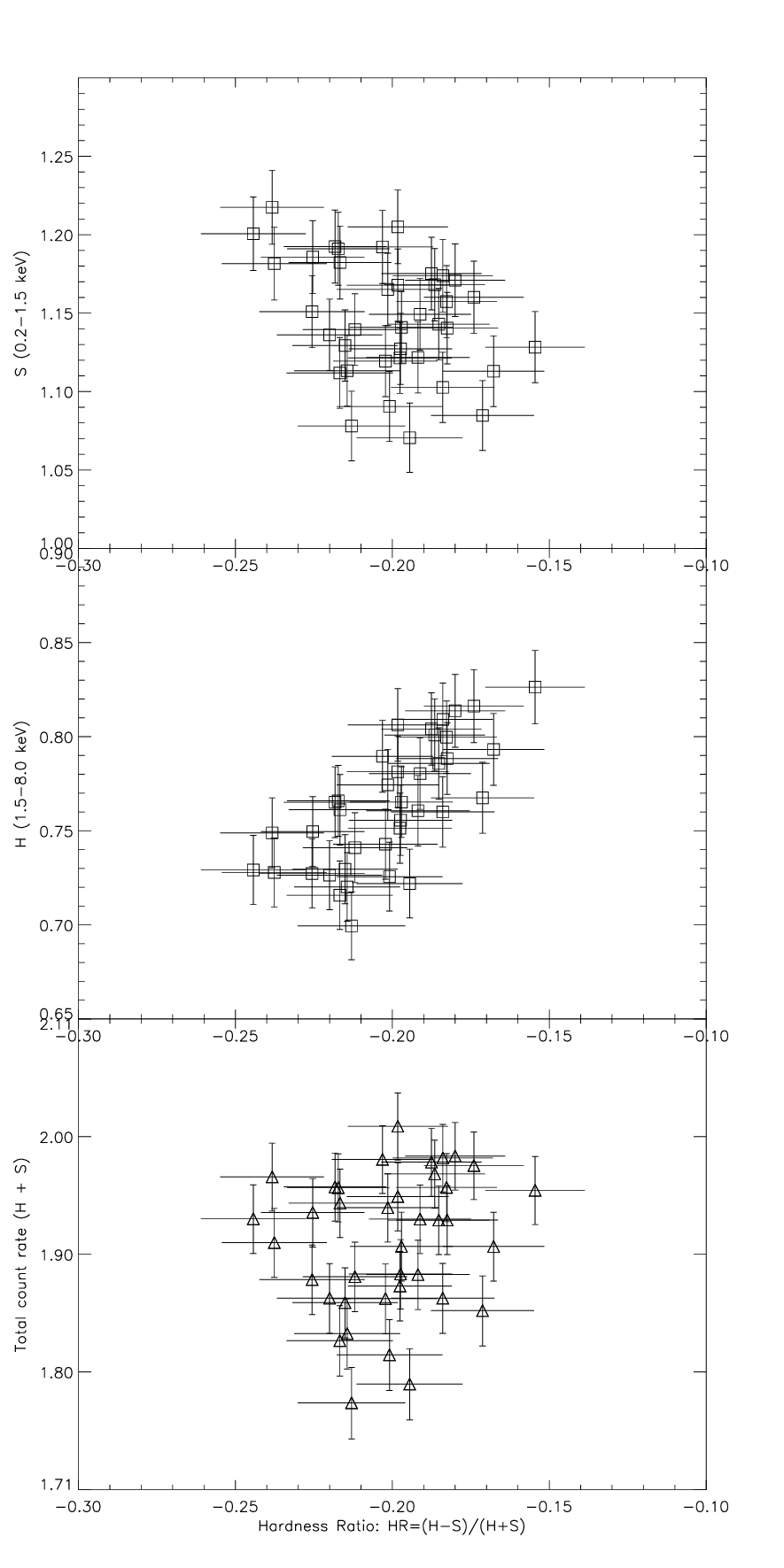,width=9cm,angle=0}}
\caption{Count rate in the two bands (S and H, used for the hardness ratio) 
vs. hardness ratio and total count rate vs. hardness ratio}
\label{colhr}
\end{figure}

\begin{figure}[!t]
\centerline{\psfig{figure=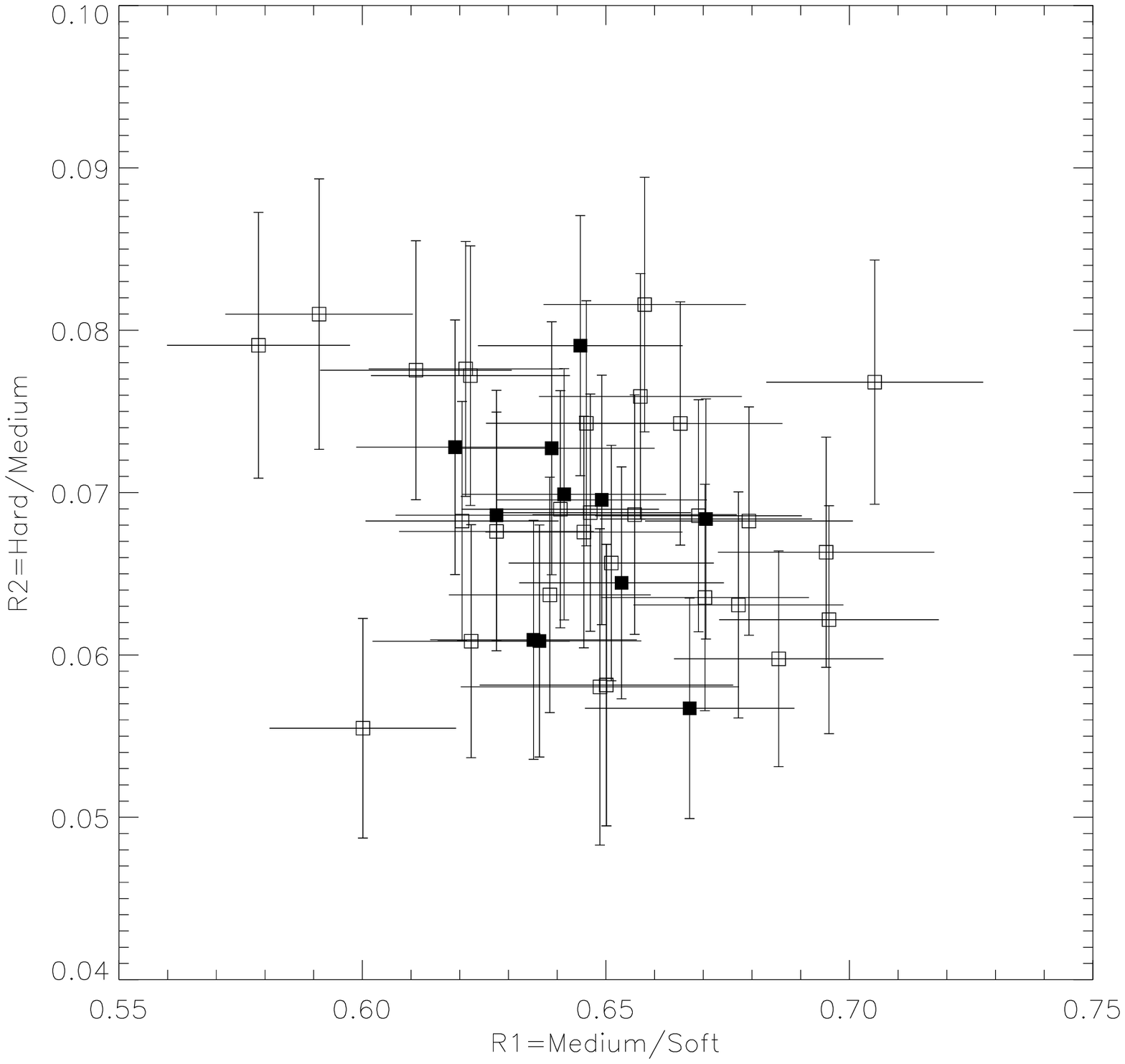,width=9cm,angle=0}}
\caption{Color-color diagram (the three energy bands are defined in
Table~\ref{tflux}. The filled points are those occurring during the low flux
regime (i.e. the dip between $5.0\times 10^4$ s and $7.5\times 10^4$ s; see
Figure~\ref{ltc}).}
\label{hr-col}
\end{figure}

The hardness ratio plotted in Figure~\ref{ltc} (Panel E) is calculated as 
$HR=\frac{Hard-Soft}{Hard+Soft}$ where Hard (H) is the number of counts in 
the 1.5-8.0 keV band and Soft (S) is the number of counts in the 0.2-1.5 keV band.
This curve strongly suggests that some spectral variability occurred during the
observation. A $\chi^2$ test against the hypothesis of constant value (equal to
the average value) gives $\chi^2/\nu=57.5/37$, with a 2\% probability that the
hardness ratio is constant. 
As suggested by Figure~\ref{ltc} and, more strongly, by Figure~\ref{colhr}, 
there is a correlation between the medium + hard band flux and the hardness 
ratio. This suggests that the variability is stronger in these bands than in the
soft one.
We find no evidence of correlations between total flux and hardness ratio
(Figure~\ref{colhr}, bottom panel), and X-ray colors with changing flux
(Figure~\ref{hr-col}). 
Colors are defined as R1= Medium/Soft and R2=Hard/Medium, with Soft, Medium and
Hard energy bands as in Table~\ref{tflux} (see also Figure~\ref{ltc}).

A power spectrum analysis performed on the photon counts with a 3.24 s
resolution in the frequency range $10^{-5}-0.3$ Hz yields a flat frequency 
distribution that 
can be fit with a single power law with a slope of $\sim4\times10^{-2}$ with
$\chi^2/\nu=64.7/75$, and is consistent within $2\sigma$ with the Poisson noise 
in the frequency range sampled by the observation.


\begin{deluxetable}{l l l l l}
\tabletypesize{\footnotesize}
\tablecaption{Variability of the nucleus of M33}
\tablewidth{0pt}
\tablehead{
 & \colhead{Full band}& \colhead{Soft band}
 &\colhead{Medium band} & \colhead{Hard band}\\
 & \colhead{0.2-8.0 keV}& \colhead{0.2-1.5 keV}
 &\colhead{1.5-4.5 keV} & \colhead{4.5-8.0 keV}}
\startdata
Avg count rate                   &$1.96\pm0.05$ c/s &$1.15\pm0.04$ c/s &$0.72\pm0.03$ c/s &$0.052\pm0.008$ c/s\\
RMS frac.var.\tablenotemark{(a)} &  2.5 \%          &  2.2\%           & 3.3\%            &\nodata \\
$\chi^2/\nu$\tablenotemark{(b)}  & 148.8/37         & 98.8/37  	       & 107.8/37         & 39.1/37  \\
KS probability\tablenotemark{(c)}&$1.1\times10^{-5}$&$5.9\times10^{-6}$&$1.3\times10^{-3}$&$0.06$\\
Timescale\tablenotemark{(b)}     &$\sim5000$ s      &$\sim5000$ s      &$\sim 5000$ s     &\nodata \\
\enddata
\tablenotetext{a}{Root Mean square fractional variation (only where higher than 
the $3 \sigma$ level)}
\tablenotetext{b}{calculated against the the hypothesis of having a constant
count rate equal to the average value.}
\tablenotetext{c}{Kolmogorov-Smirnov test probability against the hypothesis of
constant rate.}
\tablenotetext{d}{Variability timescale from the Collura test \citep{coll}}
\label{tflux} 
\end{deluxetable}

\section{SPECTRAL ANALYSIS}
\label{spec}
\subsection{Integrated spectrum}
\label{intspec}
\begin{figure}[t]
\centerline{\psfig{figure=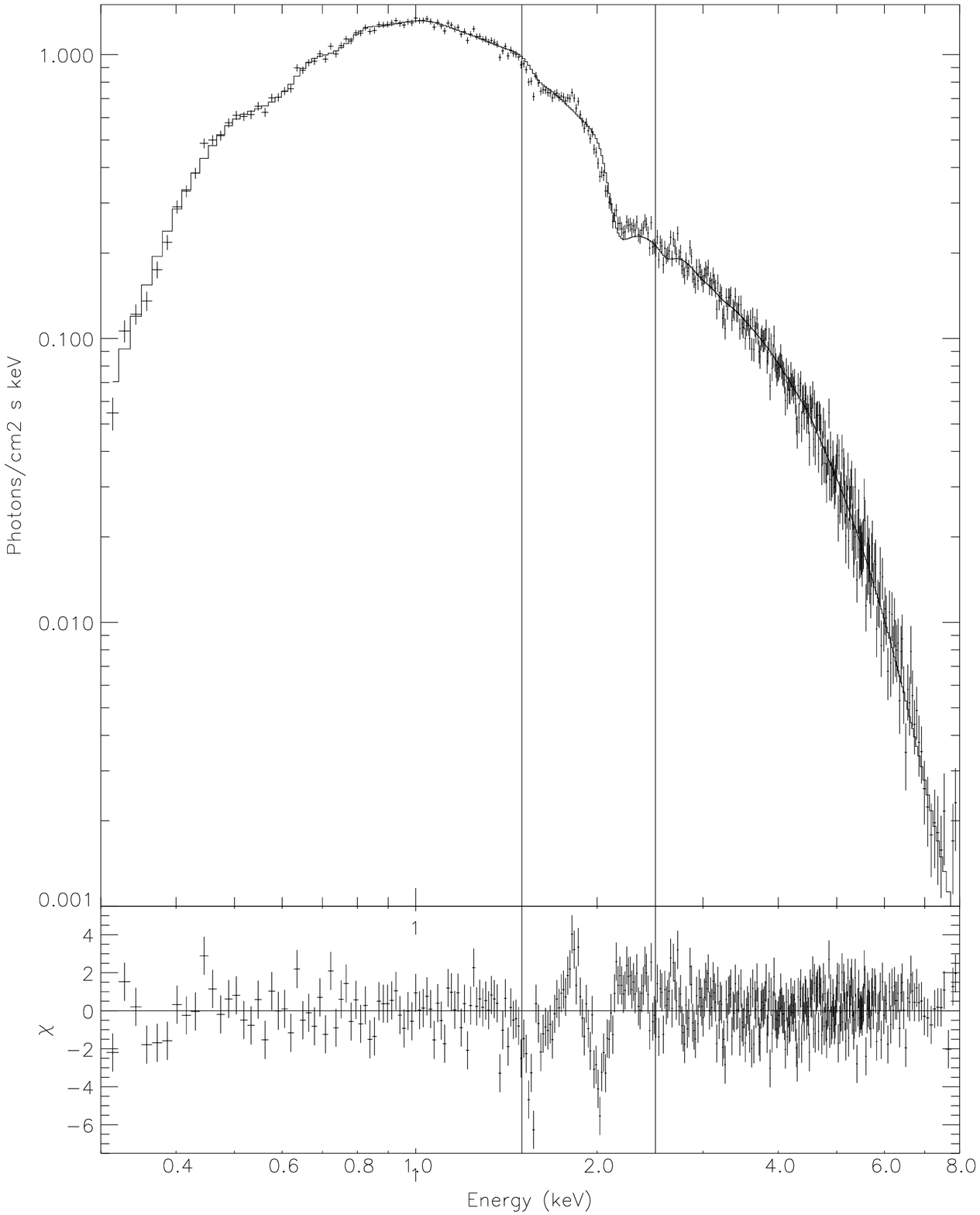,width=15cm,angle=0}}
\caption{Spectrum of M33-X8 and residuals over the best fit model
(Table~\ref{fit}). The energy channels between the two vertical lines (1.5-2.5 
keV) have been
excluded from the fitting because of an ebound conversion problem (see
Section~\ref{data}) }
\label{specf}
\end{figure}

The source spectrum appears to be affected by a PHA to PI conversion
problem\footnote{see
http://asc.harvard.edu/cal/Links/Acis/acis/Cal\_projects/index.html (note added
by Alexey Vikhlinin on 3/1/2002)} 
that causes spurious emission and absorption features between 
$\sim1.5$ and $\sim2.5$ keV. Pending resolution of this problem, the spectral 
analysis
has been conducted excluding the energy range affected by this problem. This 
left us with two data ranges: 0.3-1.5 keV and 2.5-8.0
keV, that were analyzed simultaneously. Moreover, the spectrum has been
corrected for the quantum efficiency degradation of ACIS at low energies, using
the CIAO tool {\sc corrarf}. We find that the total spectrum can be
described with a three components model: a powerlaw with $\Gamma\sim2.0$, a MCD 
with kT$_{int}\sim 1.0$ and a Raymond-Smith model (RS) with
kT$\sim0.18$ and solar abundance. The residuals, however, still show an excess 
at $\sim 1$ keV, that can be fitted with a gaussian at $\sim0.9$ keV with 
$\sigma\sim0.12$ keV. After this new component, the reduced $\chi^2$ becomes 
1.27, with an improvement of 82 over 3 additional d.o.f., yielding an F-test 
probability lower than 0.1\%. This bump could
be the signature of an overabundance of Fe and/or Ne in an environment of 
warm matter. We thus tried to model it by allowing the Fe and Ne abundance
to vary in the RS model, but the model  was not able to reproduce it:
the best result has been obtained having the Ne abundance four times higher than
the solar value, but the residuals still show a narrow gaussian-like feature at
1 keV that cannot be accounted for by varying the abundance of other elements.
Thus, the feature cannot be associated unambiguosly to any of the continuum
components. 
Figure~\ref{specf} shows the best fit model residuals, including also the range
that has been excluded from the fit. It also shows that there is no evidence of
any systematic excess in the residuals, thus the value of the reduced
$\chi^2$ can be entirely ascribed to the statistic scattering of the data.
The full set of best fit parameter values is reported in 
Table~\ref{fit}. The total flux is $F_{0.3-8.0keV}=2.35\times10^{-11}$ \flux, 
that, at the distance of M33, gives a luminosity of $1.5\times 10^{39}$ erg 
s$^{-1}$.

The flux of the
RS component is roughly consistent with the value of the 20''
extended emission reported by \citet{schul} with ROSAT/HRI. To investigate this 
possibility further, we
extracted the spectrum of the circumnuclear region of M33 X-8 from one of the
publicly archived on-axis observations of this source (Obs Id: 786, Exposure
time: 50 ksec), using the same elliptical region described in Section~\ref{data}
but excluding a circle with radius 4.92'' (10pixels) centered on the nuclear 
source (heavily piled up in this observation). The resulting spectrum 
($\sim3200$ counts) does not match the RS in either shape or intensity. It can 
be well described  by a power-law with $\Gamma=1.4$, with a flux of 
$4\times10^{-13}$ \flux in the 0.3-8.0 keV band ($L_{0.3-8.0}=2.5\times10^{37}$ 
erg/s), $\sim$1/4 of the RS flux. So, most of the RS component originates from a 
region with a radius smaller than $\sim5$'', corresponding to $\sim17$ pc at 
the distance of M33.  
\subsection{Spectral variability}
\begin{figure}[t]
\centerline{\psfig{figure=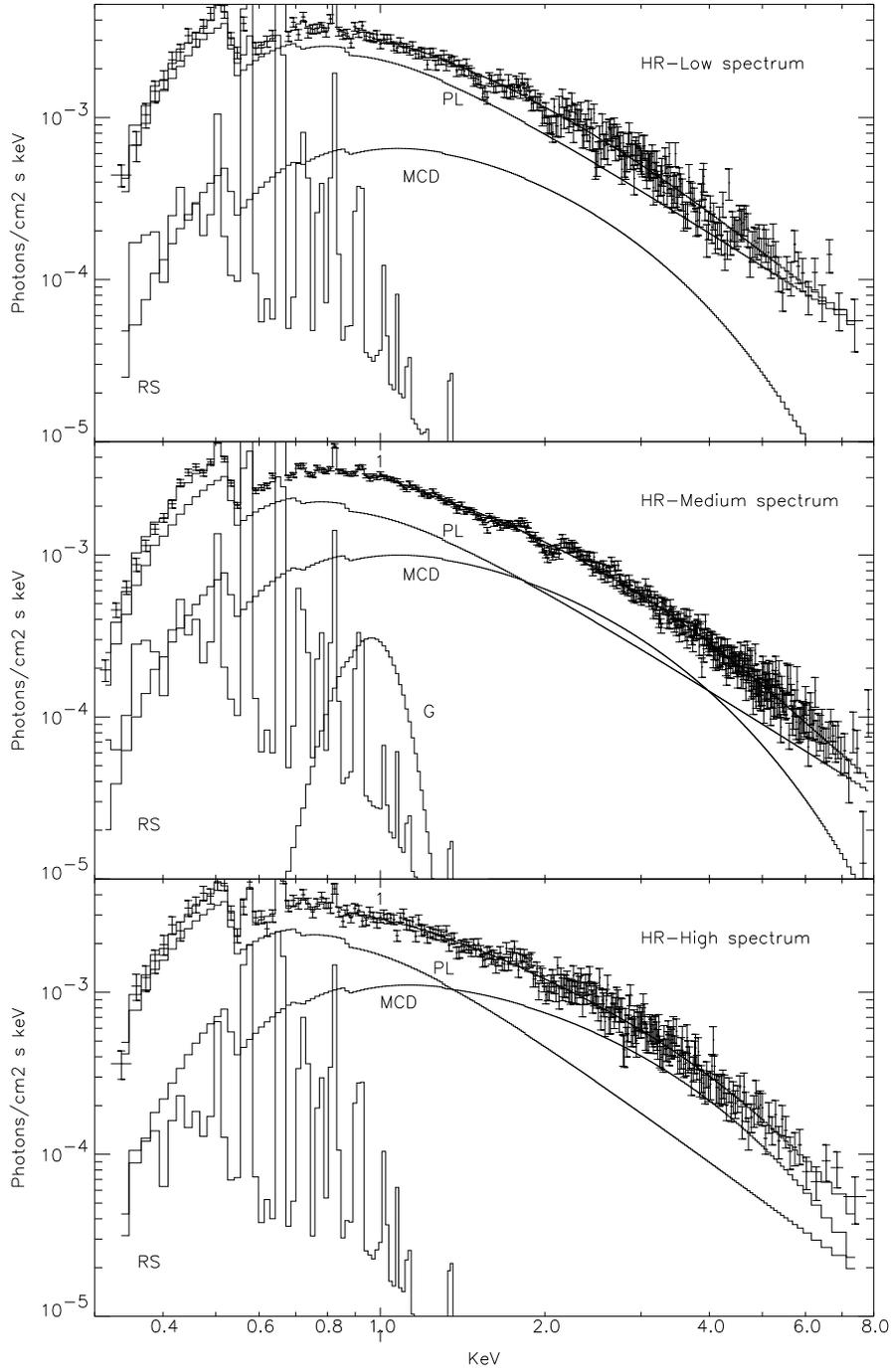,width=13cm}}
\caption{Data and best fit models (unfolded for the instrumental response) for 
the three phase-selected spectra,
deconvolved from the instrumental response. PL=Power-law; MCD=Multicolor black
body disk; RS=Raymond-Smith; G=Gaussian}
\label{comp}
\end{figure}

Figure~\ref{ltc}-E suggests spectral variability. In order to investigate
further this variability, we have divided the data into three phases, according
to the value of the hardness ratio. We define the ``{\it HR-low}'' state as the 
one where the hardness ratio (HR) is lower than -0.22, the ``{\it HR-medium}'' 
state as the one with -0.22$<$HR$<$-0.18 and the ``{\it HR-high}'' state as the 
one where HR$>$-0.18. We
have selected three spectra according to this criterion and we have analyzed
them separately, using the same spectral component as in Section~\ref{intspec}.
The results are in Table~\ref{fit}. We find that the Gaussian component is not 
needed for the {\it HR-low} and 
{\it HR-high} states. This is not surprising, as each of these spectra includes
only $\sim 10\%$ of the total counts of the source, and the gaussian component
is relatively faint. Examining the other two spectral components we find that: 
{\bf a)} the power-law photon index $\Gamma$ does not change significantly in 
the three spectra,
while the flux decreases from the {\it HR-low} to the {\it HR-medium} and 
{\it HR-high} states; 
{\bf b)} the errors of the MCD components in the {\it HR-low} state are rather
large, however there is evidence of the disk kT increasing from the medium
to the hard state; 
{\bf c)} we do not detect any variability in the RS, consistent with possibly 
extended emission from hot ISM.
Figure~\ref{comp} shows the three spectra,
deconvolved from the instrumental response, with the best fit model components.
From this figure we see a progression of larger MCD contribution (relatively to
the power-law) going from the {\it HR-low} to the {\it HR-high} spectral states.

\begin{deluxetable}{l l l l l l}
\tabletypesize{\footnotesize}
\tablecaption{Spectral Analysis - Best fit parameters}
\tablewidth{0pt}
\tablehead{
  \colhead{Component}& \colhead{Parameter}& 
  \multicolumn{4}{c}{Integrated emission\tablenotemark{(1)}}\\ 
  &&\colhead{Total spectrum} 
  & \colhead{HR-Low phase}& \colhead{Medium phase}& \colhead{HR-High phase}
}
\startdata
             &N$_H$ ($\times10^{21}$cm$^{-2})$&$2.0^{+0.2}_{-0.3}$   &$2.1^{+0.4}_{-0.3}$   &$2.0^{+0.2}_{-0.3}$   &$2.1\pm0.6$\\[0.2cm]
Power law    &$\Gamma$                        &$2.18\pm+0.07$        &$2.14^{+0.11}_{-0.09}$&$2.12^{+0.11}_{-0.07}$&$2.5^{+0.2}_{-0.3}$\\[0.2cm]
             &Flux ($10^{-11}$ \flux)         &$1.45\pm0.09$         &$1.9\pm0.2$           &$1.4\pm0.1$           &$1.3\pm0.2$\\[0.2cm]
MCD          &kT$_{int}$ (keV)                &$1.14\pm0.05$         &$1.0^{+0.3}_{-0.2}$   &$1.11^{+0.06}_{-0.07}$&$1.29\pm0.08$\\[0.2cm]
             &R$\sqrt{\cos\theta}$ (km)       &$39^{+4}_{-2}$        &$43^{+18}_{-10}$      &$42\pm4$    	   &$35^{+4}_{-5}$\\[0.2cm]
             &Flux ($10^{-12}$ \flux)         &$8.5\pm1.8$           &$5^{+4}_{-5}$         &$9\pm2$               &$11^{+4}_{-3}$\\[0.2cm]
Raymond-Smith& kT (keV)	                      &$0.18^{+0.01}_{-0.02}$&$0.19^{+0.04}_{-0.02}$&$0.17\pm0.02$         &$0.20\pm0.05$\\[0.2cm]
             &Flux ($10^{-12}$ \flux)         &$1.7\pm0.4$           &$1.7^{+1.1}_{-0.7}$   &$1.8\pm0.9$           &$1.1^{+4.4}_{-0.7}$ \\[0.2cm]
Gaussian     & E (keV)                        &$0.96^{+0.03}_{-0.10}$&\nodata		    &$0.94^{+0.04}_{-0.10}$&\nodata\\[0.2cm]
             &$\sigma$ (keV)                  &$0.10^{+0.05}_{-0.04}$&\nodata		    &$0.11^{+0.09}_{-0.04}$&\nodata\\[0.2cm]
             &Flux ($10^{-13}$ \flux)         &$1.7^{+4.3}_{-0.6}$   &\nodata		    &$2.2^{+3.3}_{-0.6}$   &\nodata\\[0.2cm] 
             &$\chi^2/\nu$                    &434/333               &210/203		    &363/314		   &182/198\\[0.2cm]
\cutinhead{Circumnuclear emission\tablenotemark{(2)}}
             &N$_H$ ($\times10^{21}$cm$^{-2})$&$0.6\pm0.3$           &\nodata   &\nodata&\nodata\\[0.2cm]
Power law    &$\Gamma$                        &$1.40\pm0.10$         &\nodata   &\nodata&\nodata\\[0.2cm]
             &Flux ($10^{-13}$ \flux)         &$4.7\pm0.5$           &\nodata   &\nodata&\nodata\\[0.2cm]
             &$\chi^2/\nu$                    &54/66                 &\nodata	&\nodata&\nodata\\[0.2cm] \hline
\enddata
\tablenotetext{(1)}{From Obs Id 2023}
\tablenotetext{(2)}{From Obs Id 786; see Section~\ref{intspec}}
\tablecomments{Unabsorbed fluxes are calculated in the 0.3-8.0 energy band.
$\theta$ is the inclination angle of the disk in the MCD model. The Gaussian 
component was not needed to fit the {\it low} and {\it high} phases. Quoted
errors are at 90\% confidence level for one interesting parameter.}
\label{fit} 
\end{deluxetable}

\section{DISCUSSION}
\label{disc}
The light curve of M33-X8 shows a significant low-amplitude variability, with a 
$\sim10\%$
variation on a timescale of a few thousands of seconds. This timescale is
particularly evident both in the full band light curve, where the source
count rate drops and suddenly goes back to the original value in
$\sim3\times10^4s$,
and in the medium band light curve, where the count rate oscillates a few times
around its average value of $\sim0.71$ c/s. This result strongly confirms the 
variability on a 3000 s timescale found by \citet{peres} in the {\it Einstein} 
data of the nucleus of M33. However, the power spectrum is flat, and does not 
evidence any periodical or quasi-periodical oscillation. 
The hardness ratio is also variable, showing evidence of spectral variability,
that appears to be correlated mostly with the emission at energy $\gtrsim 1$
keV. 

The spectral analysis reveals the presence of four components.  Two of these, a
Raymond-Smith component and a broad ``Gaussian'' line feature contribute
relatively little to the total spectrum, and do not present any compelling
evidence of variability. We will not discuss them any
further. The two main components are a strong
$\Gamma\simeq2$ power-law and a kT$=0.9-1.3$ keV multicolor disk component
(MCD). The
parameters of the power-law and of the MCD are in good agreement with the
results found by \citet{parmar}, \citet{takano}, \citet{dubus2}. However, their
contribution to the total spectrum is variable, as showed by the HR-selected
spectra (Table~\ref{fit}): the {\it HR-low} phase is characterized by a brighter
power-low and a fainter disk. While we do not find any further variation of the
power-law flux from the {\it HR-medium} to the {\it HR-high} state, the MCD flux
seems to increase with increasing kT. This result may suggest that the
variability is linked with fluctuations in the accretion rate.

For the MCD component, the lower limit of 37 km to the inner radius of the disk 
(Table~\ref{fit}; 
the limit is due to the presence of the $\sqrt{\cos\theta}$ factor accounting 
for the inclination angle of the disk) implies a lower limit to the mass of the
compact object of
$\sim4.2M_{\odot}$. The Eddington limit for this mass is 
L$_{Edd}=6.3\times 10^{38}$ erg/sec, that is consistent with the power-law + 
disk emission of 
the source. This is in agreement with the conclusion of \citet{parmar}, 
\citet{takano} and \citet{dubus2} that M33 X-8 is most likely a normal binary 
system containing a stellar mass black hole that 
happens to be at the center of the galaxy, rather than an AGN. Although not
extreme, the spectral variability of M33 X-8 is reminiscent of what has been 
observed in black hole binaries. In the
Galactic black hole binaries we usually observe a ``high/soft state'', 
dominated by a hot disk, sometimes associated with
a faint power-law tail, and a ``low/hard state'' dominated by
a power-law spectrum, with a colder disk component \citep[see, e.g., ][]{esin}. 
The soft state is explained with high accretion rate, with the accreting disk
going down toward the last stable orbit, while the hard state is associated to a
lower accretion rate that causes the disruption of the inner part of the disk  
and the presence of a comptonized corona wrapping the central source.
In the case of M33 X-8 both disk and power-law component are always present, but
the power-law dominated {\it HR-low} state corresponds with a lower flux in the medium
energy band (1.5-4.5 keV). With increasing medium band flux the spectrum 
evolves to be dominated by the disk component.

The source that resembles more M33 X-8 is probably LMC X-3 \citep{nowak}, a
black
hole candidate with a probable mass of $9 M_{\odot}$ and a luminosity of up to
$4\times 10^{38}$ erg/s. M33 X-8 presents many similarities with LMC X-3: the
thermal spectrum, that can be modelled with a MCD (with kT$\sim 0.8$ in LMC X-3
and kT$\sim 1.2$ in M33 X-8) and a steep power-law; the presence of variability 
on timescales of some ten
thousands of seconds; a long timescale periodicity (116 d for M33 X-8 and 99 or
198 d for LMC X-3, see \citealp{dubus} and \citealp{cowley}) that appears to 
vary or even disappear in a few years \citep{nowak,parmar}. Also, LMC X-3 is
characterized by a lack of fast variability (on timescales $\lesssim 1$ s), but
this feature cannot be investigated in M33 X-8, due to the lack of data with
high enough temporal resolution. However the power-law component is relatively
more prominent in M33 X-8.

Concluding, the \cha spectral variability observations of the nucleus of M33
(X-8) strongly agree with the suggestion that this source is an accreting
$\gtrsim5$ M$_{\odot}$ black hole with a variable accretion disk, comparable to
Galactic black hole binaries.

We thank M. Krauss and F. Nicastro for their help in 
solving the calibration problems, J. MacDowell for interesting discussions and
A. Kong, D. Schwartz and J. Drake for useful comments on the manuscript. 
This research has been supported by CNAA and by NASA contract NAS 8-39073 (CXC).
{}

\end{document}